\documentclass[aps,prl,twocolumn,showpacs,groupedaddress,superscriptaddress,footinbib,longbibliography]{revtex4-1}

\usepackage{natbib,hyperref}
\hypersetup{colorlinks=true, citecolor=blue, urlcolor=blue, linkcolor=blue}
\usepackage[english]{babel}
\usepackage[utf8]{inputenc}
\usepackage{graphicx}
\usepackage[caption=false]{subfig}
\usepackage{amsmath}
\usepackage{amsfonts}
\usepackage{amssymb}
\usepackage{color,soul}
\setulcolor{red}
\usepackage{easyReview}
\usepackage{xcolor}

\begin{document}


\title{Controlling vortical motion of particles in two-dimensional driven superlattices}
\author{Aritra K. Mukhopadhyay}
\email{Aritra.Mukhopadhyay@physnet.uni-hamburg.de}
\affiliation{Zentrum f\"ur Optische Quantentechnologien, Fachbereich Physik, Universit\"at Hamburg, Luruper Chaussee 149, 22761 Hamburg, Germany}
\author{Peter Schmelcher}
\email{Peter.Schmelcher@physnet.uni-hamburg.de}
\affiliation{Zentrum f\"ur Optische Quantentechnologien, Fachbereich Physik, Universit\"at Hamburg, Luruper Chaussee 149, 22761 Hamburg, Germany}
\affiliation{The Hamburg Centre for Ultrafast Imaging, Universit\"at Hamburg, Luruper Chaussee 149, 22761 Hamburg, Germany}

\date{\today}

\begin{abstract}
We demonstrate the control of vortical motion of neutral classical particles in driven superlattices. Our superlattice consists of a superposition of individual lattices whose potential depths are modulated periodically in time but with different phases. This driving scheme breaks the spatial reflection symmetries and allows an ensemble of particles to rotate with an average angular velocity. An analysis of the underlying dynamical attractors provides an efficient method to control the angular velocities of the particles by changing the driving amplitude. As a result, spatially periodic patterns of particles showing different vortical motion can be created. Possible experimental realizations include holographic optical lattice based setups for colloids or cold atoms.
\end{abstract}


\maketitle

\paragraph{Introduction.\textemdash}
Due to their experimental controllability, driven lattice potentials have become an important test bed for the exploration of non-equilibrium physical phenomena \cite{Salger_S_Directed_Transport_2009,Brown_PRA_Ratchet_Effect_2008,Dittrich_PRL_Classical_Quantum_2015}. The inherent non-linearity and tunable symmetries in these systems allow us to realize different non-equilibrium transport phenomena, the `ratchet effect' being one of them \cite{Astumian_PT_Brownian_Motors_2002,Bartussek_EL_Periodically_Rocked_1994,Cubero__Brownian_Ratchets_2016,Faucheux_PRL_Optical_Thermal_1995,Hanggi_AP_Brownian_Motors_2005,Magnasco_PRL_Forced_Thermal_1993,Prost_PRL_Asymmetric_Pumping_1994,Reichhardt_ARCMP_Ratchet_Effects_2017,Renzoni__Driven_Ratchets_2009,Reimann_PR_Brownian_Motors_2002,Hanggi_RMP_Artificial_Brownian_2009,Mukhopadhyay_PRE_Dimensional_Couplinginduced_2018}. A ratchet rectifies random particle motion into unidirectional particle transport in an unbiased non-equilibrium environment. Certain spatio-temporal symmetries of the system need to be broken in order to realize it \cite{Denisov_PR_Tunable_Transport_2014,Flach_PRL_Directed_Current_2000,Schanz_PRE_Directed_Chaotic_2005}. This leads to numerous applications across different disciplines, such as controlling the transport of atomic ensembles in ac-driven optical lattices \cite{Lebedev_PRA_Twodimensional_Rocking_2009,Schiavoni_PRL_Phase_Control_2003} both in the ultracold quantum \cite{Salger_S_Directed_Transport_2009} and classical regimes \cite{Brown_PRA_Ratchet_Effect_2008,Renzoni__Driven_Ratchets_2009}, colloidal transport in driven holographic optical lattices \cite{Arzola_PRL_Omnidirectional_Transport_2017}, particle separation based on physical properties \cite{Matthias_N_Asymmetric_Pores_2003,Mukhopadhyay_PRL_Simultaneous_Control_2018,Wambaugh_PRE_Ratchetinduced_Segregation_2002} and motion of vortices in type-II superconductors \cite{Lee_N_Reducing_Vortex_1999,Reichhardt_PRB_Reversible_Ratchet_2015,Reichhardt_PRB_Transverse_Acdriven_2016}. Due to the widespread applicability of such directed transport, there has been extensive research to control the strength and direction of the ratchet current. Setups using one dimensional (1D) driven lattices have been shown to effectively accelerate, slow down or even completely reverse the direction of transport \cite{Schanz_PRE_Directed_Chaotic_2005,Mukhopadhyay_PRE_Freezing_Accelerating_2016,Liebchen_NJP_Interactioninduced_Currentreversals_2012}. Two dimensional (2D) driven lattices on the other hand offer a higher variability in terms of transport direction and for particles to be transported parallel, orthogonal or at any arbitrary angle with respect to the direction of the driving force \cite{Arzola_PRL_Omnidirectional_Transport_2017,Mukhopadhyay_PRR_Controlling_Transport_2020,Reichhardt_PRE_Absolute_Transverse_2003}.

In contrast to 1D, the 2D ratchet setups also allow for the possibility to convert random particle motion into rotational or vortical motion leading to non-zero angular velocity of the particles. This is particularly interesting since it provides a method to realize rotational motion of neutral particles analogous to the motion of charged particles in a magnetic field. In fact, similar mechanisms have been used to generate artificial magnetic fields for exploring topological quantum states with cold neutral atoms in periodically modulated lattices \cite{Jotzu_N_Experimental_Realization_2014,Struck_PRL_Tunable_Gauge_2012}. However, the extensive research on symmetry-breaking induced directed transport in the classical regime has mostly focused on translational currents and the control of rotational currents has remained largely unexplored. The few existing setups either lead to a diffusive vortical motion over an extended space \cite{Denisov_PRL_Vortex_Translational_2008} or requires specially tailored potentials \cite{Tutu_PRE_Design_Twotooth_2011,Tutu_PRE_Robust_Unidirectional_2013} and temporally correlated colored noise \cite{Ghosh_PRE_Breaking_General_2003,Ghosh_PRL_Rotation_Asymmetric_2000}. Furthermore, due to the lack of spatial tunability of the underlying lattice potential, these setups do not allow patterns of multiple vortices in space analogous to the different spatial configurations of artificial magnetic fluxes in the quantum regime \cite{Hauke_PRL_NonAbelian_Gauge_2012}.

In this work, we address these key limitations and present a setup to realize controllable rotational motion of classical particles along closed spatial paths in driven superlattices. The individual lattices are modeled by a periodic arrangement of Gaussian potential wells whose depths can be individually modulated in a time-periodic manner. We show that modulating different wells with the same driving amplitude but different driving phases allow us to break the relevant symmetries and generate non-zero average angular velocities for an ensemble of particles. The angular velocities of individual trajectories can be controlled by varying the driving amplitude. Additionally, we demonstrate periodic spatial arrangements of different types of rotational motion by modulating the different potential wells with different driving amplitudes and phases.


\paragraph{Setup.\textemdash} 
We consider $N$ non-interacting classical particles of mass $m$ in a 2D potential landscape $V\left({\bf r}\equiv(x,y,0),t\right)$=$\sum_{m,n=-\infty}^{+\infty} \tilde{U}_{mn}(t) e^{-\beta\left( {\bf r} - {\bf r}_{mn}\right)^2}$ formed by a lattice of 2D Gaussian wells centered at positions ${\bf r}_{mn}=(mL,nL,0)$, $m,n$ $\in\mathbb{Z}$. The depths of the wells are modulated periodically in time by the site-dependent driving law $\tilde{U}_{mn}(t)=\tilde{V}_{mn}\left(\cos(\omega t +\phi_{mn})-1\right)$ with driving frequency $\omega$, driving amplitude $\tilde{V}_{mn}$ and a temporal phase shift $\phi_{mn}$. Introducing dimensionless variables ${\bf r}'=\frac{{\bf r}}{L}$ and $t'=\omega t$ and dropping the primes for simplicity, the equation of motion for a single particle at position  ${\bf r}=(x,y,0)$ with velocity ${\bf \dot{r}}=(\dot{x},\dot{y},0)$ reads
\begin{equation}
\ddot{\bf r}+\gamma \dot{\bf r} =\sum_{m,n=-\infty}^{+\infty} 2 \alpha U_{mn}(t) \left( {\bf r} - {\bf R}_{mn} \right) e^{-\alpha({\bf r} - {\bf R}_{mn})^2}  + \pmb{\xi} (t)\label{eqm}
\end{equation}
where $U_{mn}(t)=V_{mn}\left(\cos(t +\phi_{mn})-1\right)$ is the effective site dependent driving law with time period $T=2\pi$ and driving amplitude $V_{mn}=\frac{\tilde{V}_{mn}}{m\omega ^2 L^2}$. ${\bf R}_{mn}=(m,n,0)$ denotes the positions of the Gaussian wells, $\gamma=\frac{\tilde{\gamma}}{m\omega}$ is the effective dissipation coefficient and the parameter $\alpha=\beta L^2$ is a measure of the widths of the wells. $\pmb{\xi} (t)=(\xi_x,\xi_y,0)$ denotes thermal fluctuations modeled by Gaussian white noise of zero mean with the property $\langle \xi_i (t) \xi_j (t')\rangle = 2 D\delta_{ij}\delta (t-t')$ where $i,j \in {x,y}$ and $D=\frac{\tilde{\gamma} k_B \mathcal{T}}{m\omega ^2 L^2}$ is the dimensionless noise strength with $\mathcal{T}$ and $k_B$ denoting the temperature and Boltzmann constant respectively. Unless mentioned otherwise, we choose $V_{mn}=V$ for all the wells, $\alpha=3$ and $\gamma=0.1$. The set of all wells arranged periodically in space with a specific value of the driving phase $\phi_{mn}$ forms a sublattice of our system. Our setup is hence a driven superlattice formed by the superposition of different sublattices, each driven with a distinct driving phase $\phi_{mn}$. Possible experimental realizations of such a 2D potential include holographic optical lattices \cite{Barredo_S_Atombyatom_Assembler_2016,Kim_NC_Situ_Singleatom_2016,Nogrette_PRX_SingleAtom_Trapping_2014,Stuart_NJP_Singleatom_Trapping_2018,Arzola_PRL_Omnidirectional_Transport_2017} or optical superlattices \cite{Lohse_NP_Thouless_Quantum_2016} with the lattice depth modulated via standard amplitude modulation techniques \cite{Alberti_NJP_Atomic_Wave_2010,Arnal_PRA_Evidence_Cooling_2019}. The rotational dynamics of particles in such a setup could be observed with colloidal particles or with cold atoms in the classically describable regime of microkelvin temperatures \cite{Arzola_PRL_Omnidirectional_Transport_2017,Renzoni__Driven_Ratchets_2009}.

\begin{figure} 
	\centerline{\includegraphics[scale=0.35]{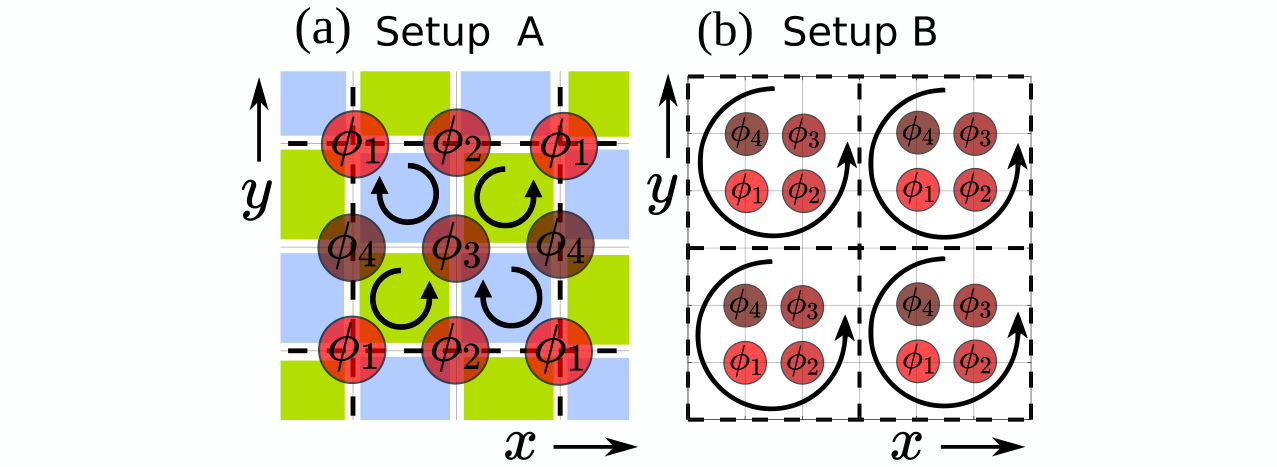}}
	\caption{Schematic representation of the two superlattice setups A and B formed by the superposition of four square sublattices driven with an amplitude $V$ but at different phases $\phi_i=\frac{(i-1)\pi}{2}$, $i=1,2,3,4$. Each colored (red) circle denotes the position of an individual Gaussian well. The thick dashed lines in black denote the boundary of the lattice unitcells. The spatial period of setup A is $(2,2,0)$ whereas that of setup B is $(3,3,0)$ due to the presence of empty sites without any wells. The blue and green regions in Fig.~(a) denote plaquettes having clockwise and anti-clockwise chirality with respect to the spatial orientation of the wells with driving phases $\phi_i$. Remaining parameters are: $V=0.41$, $\alpha=3$, $\gamma=0.1$.}\label{fig1}
\end{figure}

\paragraph{Rotational current due to symmetry breaking.\textemdash} 
The asymptotic dynamics of particles in our setup can be either confined within a lattice unitcell such as in linear oscillatory motion or vortical motion along arbitrary closed spatial curves. There can also be unconfined diffusive or ballistic motion throughout the lattice. Different particles exhibiting vortical motion can, in general, possess different angular velocities. Hence in order to distinguish vortical motion of a trajectory from ballistic, diffusive and vortical dynamics of other trajectories, we use the angular velocity ${\bf \Omega}(t) = \left[{\bf \dot{r}}(t) \times {\bf \ddot{r}}(t)\right]/{\bf \dot{r}}^2(t)$ which is equivalent to the definition of curvature of planar curves measuring the speed of rotation of the velocity vector about the origin \cite{Denisov_PRL_Vortex_Translational_2008,Stoker__Differential_Geometry_1989}. Since the particle dynamics is confined to the $xy$ plane, the only possible non-zero component of ${\bf \Omega}(t)$ is along ${\bf \hat{z}}$, the unit vector along the $z$ direction. The mean angular velocity of a trajectory is defined as ${\bf \bar{\Omega}}=\frac{1}{t}\lim\limits_{t\rightarrow \infty} \int_{0}^{t} {\bf \Omega}(t') dt'$. For trajectories rotating along a closed spatial curve with period $\eta T$, the mean angular velocity can be expressed as ${\bf \bar{\Omega}}=\frac{2\pi \tau}{\eta T} {\bf \hat{z}}=\frac{\tau}{\eta} {\bf \hat{z}}$ (since $T=2\pi$), where $2\pi \tau$ denotes the \textit{total curvature} of the curve with the \textit{turning number} $\tau$ defined as the number of times the velocity vector winds about its origin \cite{Berger__Differential_Geometry_1988}. The net rotational current, defined as the mean angular velocity of an ensemble of particles with different initial conditions, is given by ${\bf J_{\Omega}}=\langle {\bf \bar{\Omega}} \rangle$ where $\langle ... \rangle$ denotes the average over all trajectories. Since the only possible non-zero components of ${\bf \Omega}(t),{\bf \bar{\Omega}}$ and ${\bf J_{\Omega}}$ is along ${\bf \hat{z}}$, we drop the symbol ${\bf \hat{z}}$ henceforth.

The necessary condition for any setup to exhibit a net rotational current is to break the symmetries which keeps the system invariant but changes the sign of the angular velocity ${\bf \Omega}(t)$ \cite{Denisov_PRL_Vortex_Translational_2008}. There are only two symmetry transformations which can change the sign of ${\bf \Omega}(t)$: (i) time reversal together with optional spatial inversion and space-time translations: $S_{t}$: $t\longrightarrow -t + t'$, $\mathbf{r}\longrightarrow \pm\mathbf{r} + \pmb{\delta}$ and (ii) parity or reflection $\mathcal{P}$ about any plane perpendicular to the $xy$ plane with optional spatial rotation $\mathcal{R}$ in the $xy$ plane and space-time translations: $S_{p}$: $\mathbf{r}\longrightarrow \mathcal{R}\left(\mathcal{P}\mathbf{r}\right) + \pmb{\delta}$, $t\longrightarrow t + t'$. Since our setup is dissipative, $S_{t}$ is broken independent of our choice of the lattice potential $V\left({\bf r},t\right)$. However, the superlattice potential allows us to preserve or break the symmetry $S_{p}$ by controlling the driving phases of the underlying sublattices. In order to illustrate this, we consider two setups A and B (Figs.~\ref{fig1}(a,b)) each consisting of four square sublattices with the same driving amplitude $V=0.41$ but different phases $\phi_{i}=\frac{(i-1)\pi}{2}$, $i=1,2,3,4$. The sublattices in setup A have lattice vectors $(2,0,0)$ and $(0,2,0)$, hence the setup has a spatial period ${\bf L_A}=(2,2,0)$. In contrast, the setup B has a spatial period ${\bf L_B}=(3,3,0)$ with the lattice vectors being $(3,0,0)$ and $(0,3,0)$. As shown in Fig.~\ref{fig1}(a), the arrangement of the sublattices allows us to consider the unitcell of the setup A as a collection of four distinct spatial domains or plaquettes. The plaquettes are characterized by clockwise or counter-clockwise arrangement of Gaussian wells with driving phase $\phi_{i}$, i.e. of opposite chirality. Since the parity transformation $S_{p}$ reverses chirality, each of these plaquettes break the $S_{p}$ symmetry. However since the unitcell has equal number of plaquettes with opposite chirality (two clockwise and two anti-clockwise), the unitcell and hence the entire setup A is symmetric with respect to $S_{p}$. This implies that although the setup A might allow trajectories with different mean angular velocities ${\bf \bar{\Omega}}$, the net rotational current ${\bf J_{\Omega}}$ must be zero. In contrast, the entire unitcell of setup B has an anti-clockwise chirality which can be reversed by $S_{p}$ and hence the setup B breaks $S_{p}$ symmetry. As a result one can expect ${\bf J_{\Omega}}$ to be non-zero. 



\begin{figure}[t]
	\centerline{\includegraphics[scale=0.3]{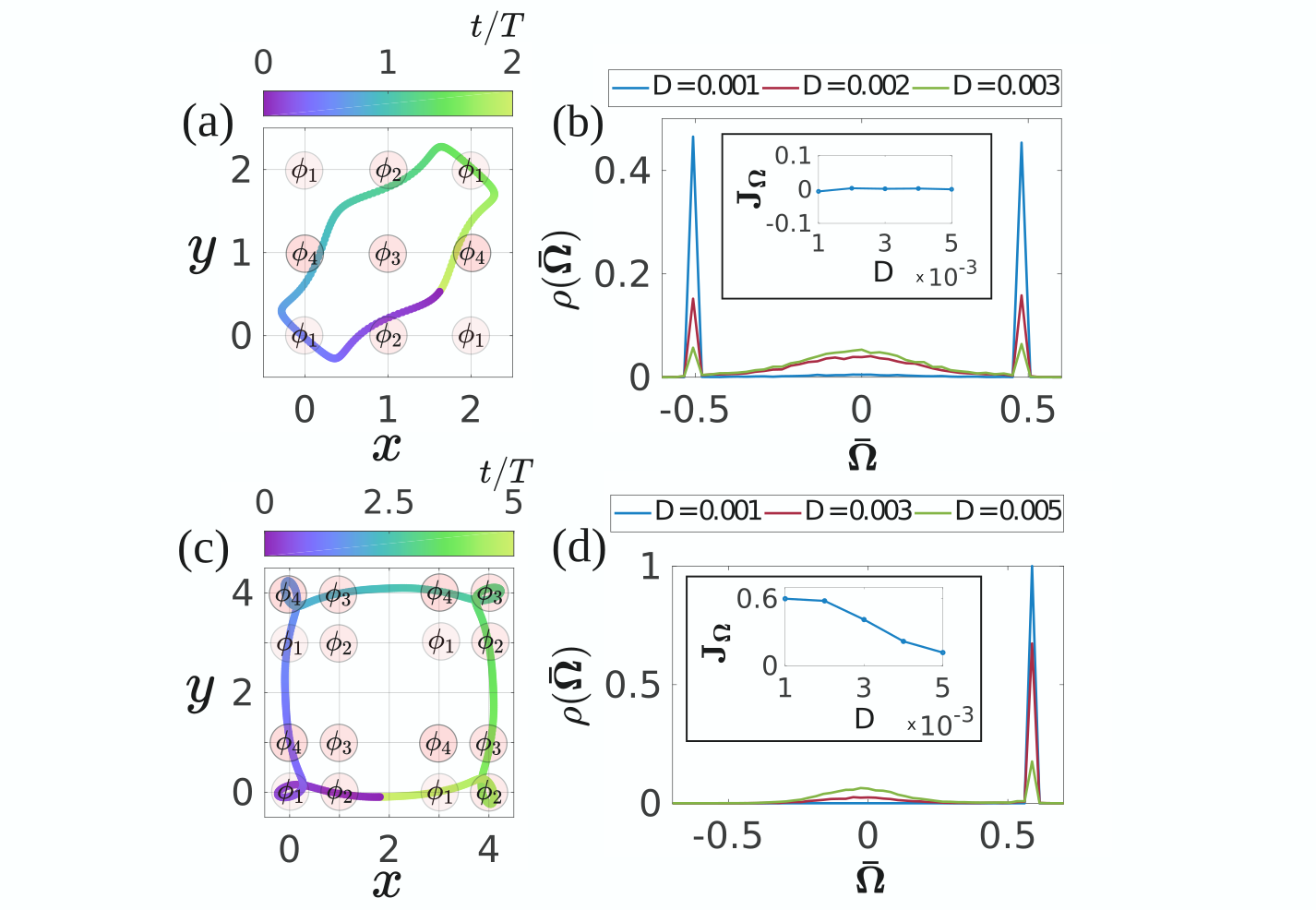}}
	\caption{Typical trajectories exhibiting rotational motion in (a) setup A and (c) setup B respectively over one time period of rotation (in colorbars). The colored circles denote the positions of individual Gaussian wells with different driving phases $\phi_i$. Figures (b) and (d) show the fraction of particles $\rho({\bf \bar{\Omega}})$ possessing mean angular momentum ${\bf \bar{\Omega}}$ for different noise strengths $D$ in setup A and B respectively. The insets show the variation of the net rotational current ${\bf J_{\Omega}}$ with $D$. Remaining parameters are the same as in Fig.~\ref{fig1}.}\label{fig2}
\end{figure}

In order to verify our symmetry analysis and explore the behavior of rotational current in our system, we initialize $N=10^4$ particles randomly within a square region $x,y \in [-100,100]\times [-100,100]$ in both setups A and B with small random velocities $v_x,v_y \in [-0.1,0.1]$. Subsequently we time evolve our ensemble up to time $t_f= 10^4 T$ by numerical integration of Eq.~\ref{eqm} for different noise strength $D$. In the deterministic limit $D=0$, all the particles in setup A exhibit only rotational motion along closed curves either with mean angular momentum ${\bf \bar{\Omega}}=\frac{1}{2} $ (vortex) or $-\frac{1}{2} $ (antivortex). Fig.~\ref{fig2}(a) shows a typical trajectory in this setup having ${\bf \bar{\Omega}}=-\frac{1}{2} $. The velocity vector winds around its origin in clockwise direction once during the period of rotation $2T$, hence $\tau=-1$ and $\eta=2$. The vortical motion persists as the noise strength is increased to $D=0.001$. However most importantly, there exists an equal number of trajectories possessing ${\bf \bar{\Omega}}=-\frac{1}{2} $ and ${\bf \bar{\Omega}}=\frac{1}{2} $ signifying that the net rotational current ${\bf J_{\Omega}}=0$ (Fig.~\ref{fig2}(b)), as predicted by our symmetry analysis. Even for higher noise strength up to $D=0.003$, such a symmetry related cancellation of vortex-antivortex pairs with equal and opposite angular momentum persists, leading to a zero net rotational current. Beyond $D>0.003$, the vortical motion is destroyed resulting in a symmetric distribution of particles around ${\bf \bar{\Omega}}=0$ and hence ${\bf J_{\Omega}}=0$. The particles in setup B also exhibit rotational motion, however unlike in setup A, all the particles in setup B possess a mean angular momentum ${\bf \bar{\Omega}}=\frac{3}{5} =0.6$. An example trajectory in setup B in the deterministic limit can be seen in Fig.~\ref{fig2}(c). The velocity vector makes four anti-clockwise (at the four corners of the curve) and one clockwise (corresponding to one full rotation along the curve) winding around its origin during one period of rotation $5T$, hence $\tau=3$ and $\eta=5$. For $D\leqslant0.002$, the vortical motion is quite stable and almost all the particles in the setup rotate with ${\bf \bar{\Omega}}=0.6$ resulting in ${\bf J_{\Omega}}=0.6 $ (Fig.~\ref{fig2}(d)) in accordance with our symmetry analysis. For $D>0.002$, the particles perform diffusive motion through the lattice and the vortical motion is gradually destroyed thus decreasing the value of ${\bf J_{\Omega}}$.

\begin{figure}
	\centerline{\includegraphics[scale=0.28]{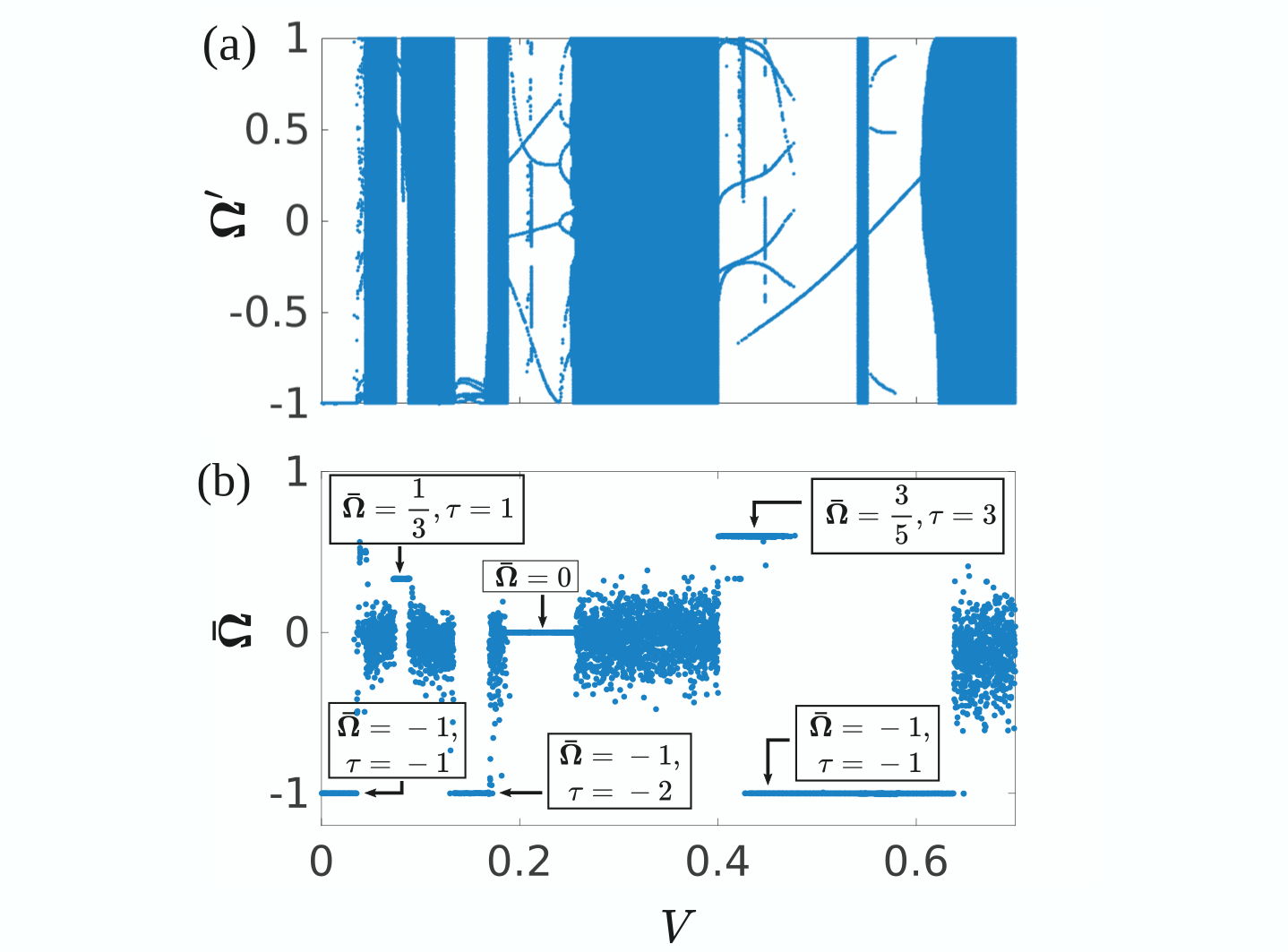}}
	\caption{(a) Bifurcation diagram of ${\bf \Omega'}(t)$ as a function of the driving amplitude $V$ depicting the chaotic (broad blue bands) and regular (thin blue lines) attractors of the setup B (see Fig.~\ref{fig1}(b)). (b) The mean angular momentum ${\bf \bar{\Omega}}$ of the attractors in Fig.~\ref{fig3}(a) as a function of $V$. The values of ${\bf \bar{\Omega}}$ for the regular attractors denoting rotational motion and the turning number $\tau$ of the corresponding closed curves are labeled with arrows. Remaining parameters are the same as in Fig.~\ref{fig1}(b).}\label{fig3}
\end{figure}

\paragraph{Control of rotational current.\textemdash} 
The question that naturally arises is that once we design a driven superlattice which breaks the $S_{p}$ symmetry, for e.g. our setup B, can we predict the value of ${\bf J_{\Omega}}$ apriori? Specifically, how does the mean angular momentum ${\bf \bar{\Omega}}$ of the trajectories depend on the system parameters? For a driven dissipative non-linear system like the present one, this can be answered by analyzing the asymptotic $t\rightarrow \infty$ particle dynamics in the deterministic limit $D=0$. The asymptotic dynamics of the particles is governed by the set of attractors underlying the phase space of the system, which can be of two types: (i) \textit{regular} attractors denoting ballistic, linear oscillatory and rotational motions (ii) \textit{chaotic} attractors denoting diffusive motion. In order to distinguish between attractors corresponding to rotational motion as compared to the others, we introduce a slightly modified angular momentum vector ${\bf \Omega'}(t) = \left[{\bf \dot{r}}(t) \times {\bf \ddot{r}}(t)\right]/\left[|{\bf \dot{r}}(t)||{\bf \ddot{r}}(t)|\right]$. Note that ${\bf \Omega'}(t)=\sin \vartheta (t) \hspace{0.1cm} {\bf \hat{z}}$ where $\vartheta (t)$ denotes the instantaneous angle between the velocity and acceleration vectors of the particle. ${\bf \Omega'}(t)$ transforms under $S_{p}$ and $S_{t}$ in exactly the same way as ${\bf \Omega}(t)$. However since the values of ${\bf \Omega'}(t)$ are bounded in the interval $[-1,1]$, as opposed to ${\bf \Omega}(t)$ which becomes large for small values of ${\bf \dot{r}}(t)$, it is a good quantity to differentiate between chaotic and regular rotational dynamics of particles. To illustrate this, we inspect the \textit{bifurcation diagram} of ${\bf \Omega'}(t)$ in Fig.~\ref{fig3}(a) as a function of the driving amplitude $V$ for our setup B by initializing particles with random position and velocities and stroboscopically monitoring ${\bf \Omega'}(t)$ after an initial transient \cite{Tabor__Chaos_Integrability_1989}. For certain ranges of values of $V$, all the particles in the setup exhibit chaotic motion (broad blue bands in Fig.~\ref{fig3}(a)) such that ${\bf \Omega'}(t)$ takes all possible values in the range $[-1,1]$. For all other values of $V$, they perform regular periodic motion resulting in only specific values of ${\bf \Omega'}(t)$. Most of these periodic motions correspond to particles performing rotational motion with different non-zero ${\bf \bar{\Omega}}$ (except for $0.19\lesssim V\lesssim 0.25$) depending on the value of $V$ as shown in Fig.~\ref{fig3}(b). This provides an efficient method to design and control the angular momentum of the trajectories in our setup by simply choosing the desired driving amplitude $V$. Our previous results (see Figs.~\ref{fig2}(c,d)) is such an example for the setup B with $V=0.41$.


\begin{figure}
	\centerline{\includegraphics[scale=0.25]{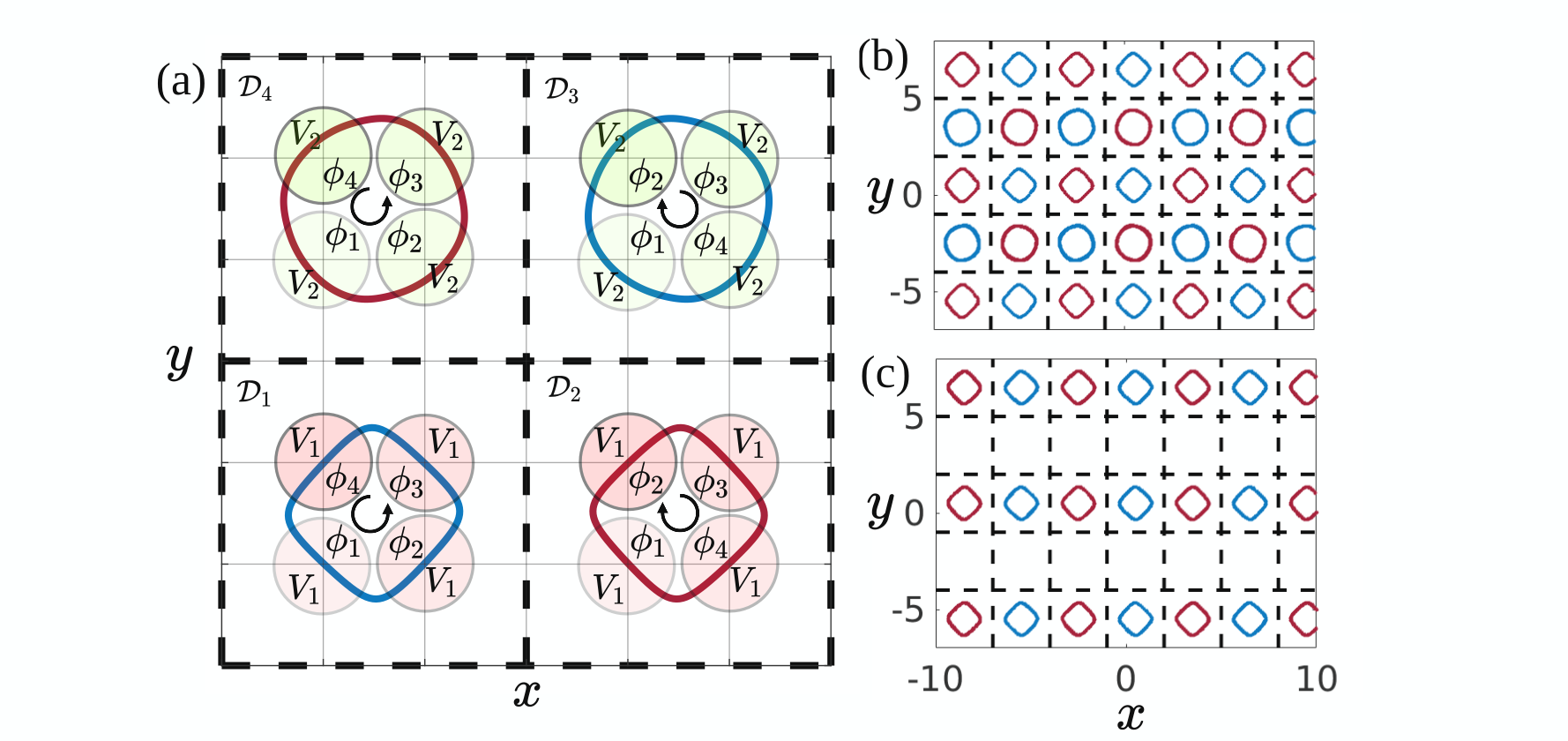}}
	\caption{(a) Schematic representation of one unitcell of our setup consisting of four plaquettes $\mathcal{D}_1$, $\mathcal{D}_2$, $\mathcal{D}_3$ and $\mathcal{D}_4$ with the thick dashed lines denoting the plaquette boundaries. The color filled circles denote the positions of individual Gaussian wells driven with amplitudes $V_1=0.51$ or $V_2=0.078$ and phases $\phi_i$. $\mathcal{D}_1$ and $\mathcal{D}_4$ ($\mathcal{D}_2$ and $\mathcal{D}_3$) have anti-clockwise (clockwise) chirality with respect to the spatial orientation of the wells with driving phases $\phi_i$. Trajectories of particles exhibiting vortical motion for $D=0$ with positive (red) and negative (blue) ${\bf \bar{\Omega}}$ have been superimposed on the unitcell. The trajectories in $\mathcal{D}_1$, $\mathcal{D}_2$, $\mathcal{D}_3$ and $\mathcal{D}_4$ have ${\bf \bar{\Omega}}=-1$, $1$, $-\frac{1}{3}$ and $\frac{1}{3}$ respectively. An extract of the spatial arrangements of the trajectories exhibiting vortical motion within different plaquettes for $D=10^{-4}$ and $D=10^{-3}$ is shown in (b) and (c) respectively. Remaining parameters are the same as in Fig.~\ref{fig1}.}\label{fig4}
\end{figure}

\paragraph{Multiple vortices.\textemdash} 
The ability to control the angular momentum of the particles with different driving amplitude $V$ allows us to design lattices with spatially periodic arrangements of multiple vortices. In order to illustrate this, we consider a specific setup as shown in Fig.~\ref{fig4}(a). It is designed such that the unitcell consists of a collection of four plaquettes $\mathcal{D}_1$, $\mathcal{D}_2$, $\mathcal{D}_3$ and $\mathcal{D}_4$. Each plaquette consists of four Gaussian wells driven at different phases $\phi_{i}=\frac{(i-1)\pi}{2}$, $i=1,2,3,4$. The plaquettes $\mathcal{D}_1$ and $\mathcal{D}_4$ possess an anti-clockwise chirality whereas $\mathcal{D}_2$ and $\mathcal{D}_3$ have clockwise chirality with respect to the spatial arrangement of the wells with driving phases $\phi_{i}$. Additionally, the wells in $\mathcal{D}_1$ and $\mathcal{D}_2$ are driven with amplitude $V_1=0.51$ and those in $\mathcal{D}_3$ and $\mathcal{D}_4$ with $V_2=0.078$. Note that these specific values of driving amplitude are chosen by consulting the bifurcation diagram in Fig.~\ref{fig3}, so as to allow only vortex trajectories having specific angular momenta. We initialize $N=10^4$ particles randomly in this setup within a square region $x,y \in [-50,50]\times [-50,50]$ with small random velocities $v_x,v_y \in [-0.1,0.1]$ and propagate the ensemble up to time $t_f= 10^4 T$. For $D=0$, the particles exhibit vortical motion at long timescales with their angular momentum being governed by the plaquette they are trapped within as shown in Fig.~\ref{fig4}(a). The particles in $\mathcal{D}_1$ and $\mathcal{D}_4$ rotate with ${\bf \bar{\Omega}}=-1$ and ${\bf \bar{\Omega}}=\frac{1}{3}$ respectively, as predicted by Fig.~\ref{fig3}(b). Note that the plaquettes $\mathcal{D}_2$ and $\mathcal{D}_3$ can be obtained by a spatial parity transformation on $\mathcal{D}_1$ and $\mathcal{D}_4$ respectively. Hence the mean angular momentum of the particles in $\mathcal{D}_2$ and $\mathcal{D}_3$ has an opposite sign as compared to the particles in $\mathcal{D}_1$ and $\mathcal{D}_4$ respectively. Even for $D=10^{-4}$, such rotational motion persists and we obtain a periodic arrangement of particles in space rotating with different angular momenta (Fig.~\ref{fig4}(b)). For a  higher strength $D=10^{-3}$, the vortical motion of particles with ${\bf \bar{\Omega}}=\pm \frac{1}{3}$ is destroyed and only the ones with ${\bf \bar{\Omega}}=\pm 1$ remain, yielding a different periodic arrangement (Fig.~\ref{fig4}(c)). Noise strengths $D\geqslant4\times 10^{-3}$ eventually destroy all the vortex trajectories.

\paragraph{Conclusions.\textemdash} 
We have demonstrated that superlatices of periodically driven localized wells provide highly controllable setups to realize different patterns of rotational motion of particles. The spatial arrangement of the lattices is responsible for breaking the relevant symmetries, thus allowing for the non-zero average angular momentum of an ensemble of particles. Our analysis of the underlying non-linear dynamical attractors provide an efficient method to control the angular momentum of the particles as well as create a variety of periodic arrangements of vortical motion with different angular momenta. Future perspectives include investigation of rotational dynamics of particles operating in the purely Hamiltonian regime without dissipation, as well as in the quantum regime with the possibility to realize spatially varying artificial magnetic fluxes.

\begin{acknowledgments}
 A.K.M acknowledges a doctoral research grant (Funding ID: 57129429) by the Deutscher Akademischer Austauschdienst (DAAD) and thanks J. Chen for insightful discussions.
\end{acknowledgments}

\bibliographystyle{apsrev4-1}
\bibliography{mybib}

\end{document}